\documentclass{aa}

\makeatletter

\renewcommand*\aa@pageof{, page \thepage{} of \pageref*{LastPage}}

\usepackage{natbib}
\bibpunct{(}{)}{;}{a}{}{,}%
\def\bibfont{\aa@bibliographyfont}%
\providecommand{\@LN}[2]{}

\makeatother

\usepackage{graphicx}
\usepackage{newtxtext}
\usepackage[smallerops,cmbraces]{newtxmath}
\usepackage{xcolor}
\definecolor{xlinkcolor}{cmyk}{1,1,0,0}
\usepackage[
bookmarks=true,         
pdfnewwindow=true,      
colorlinks=true,        
linkcolor=xlinkcolor,   
citecolor=xlinkcolor,   
filecolor=xlinkcolor,   
urlcolor=xlinkcolor,    
final=true,
]{hyperref}

\usepackage{bm} 
\usepackage{textgreek} 

\usepackage[capitalize]{cleveref} 
\crefname{section}{Sect.}{Sects.}
\creflabelformat{equation}{#2#1#3} 
\crefname{enumi}{item}{items} 

\usepackage{siunitx} 
\DeclareSIUnit[number-unit-product = ]\percent{\char`\%} 

\usepackage{csquotes} 
\definecolor{blackberry}{HTML}{8D1D75}

\usepackage[normalem]{ulem} 
\newcommand*{\code}[1]{\texttt{#1}} 
\newcommand*{\name}[1]{\textsc{#1}} 

\newcommand*{\ngc}[1]{\object{NGC\,#1}}
\newcommand*{\m}[1]{\object{M\,#1}}

\newcommand*{\gadget}[1]{\name{Gadget-#1}}

\DeclareSIUnit\parsec{pc}
\DeclareSIUnit\dex{dex}
\DeclareSIUnit\h{\mathnormal{h}}
\DeclareSIUnit\year{yr}
\DeclareSIUnit\years{yrs}
\DeclareSIUnit\arcsec{arcsec}
\DeclareSIUnit\arcmin{arcmin}
\DeclareSIUnit\Msun{M_\odot}
\DeclareSIUnit\Rsun{R_\odot}
\DeclareSIUnit\Lsun{L_\odot}
\DeclareSIUnit\Mvir{\mathnormal{M}_\mathrm{vir}}
\DeclareSIUnit\Rvir{\mathnormal{R}_\mathrm{vir}}
\DeclareSIUnit\Rhalf{\mathnormal{R}_{1/2}}
\DeclareSIUnit\erg{erg}
\DeclareSIUnit\angstrom{\text{Å}}

\newcommand*{\Msun}{\ensuremath{\mathrm{M}_\odot}} 
\newcommand*{\Rsun}{\ensuremath{\mathrm{R}_\odot}} 
\newcommand*{\Lsun}{\ensuremath{\mathrm{L}_\odot}} 
\newcommand*{\Mvir}{\ensuremath{M_\mathrm{vir}}} 
\newcommand*{\Rvir}{\ensuremath{R_\mathrm{vir}}} 
\newcommand*{\Rhalf}{\ensuremath{R_{1/2}}} 

\begin{document}

\title{Galaxy archaeology for wet mergers: Globular cluster age distributions in the Milky Way and nearby galaxies}
\titlerunning{GCs in wet mergers}

\author{
    Lucas M. Valenzuela\inst{\ref{inst:usm}}
    \and
    Rhea-Silvia Remus\inst{\ref{inst:usm}}
    \and
    Madeleine McKenzie\inst{\ref{inst:anu}}
    \and
    Duncan A. Forbes\inst{\ref{inst:cas}}
}
\authorrunning{L. M. Valenzuela et al.}

\institute{
    Universitäts-Sternwarte, Fakultät für Physik, Ludwig-Maximilians-Universität München, Scheinerstr. 1, 81679 München, Germany\label{inst:usm}\\
    \email{lval@usm.lmu.de}
    \and
    Research School of Astronomy \& Astrophysics, Australian National University, Canberra, ACT 2611, Australia\label{inst:anu}
    \and
    Centre for Astrophysics \& Supercomputing, Swinburne University, Hawthorn, VIC 3122, Australia\label{inst:cas}
}

\date{Received XX Month, 20XX / Accepted XX Month, 20XX}

\abstract
{Identifying past wet merger activity in galaxies has been a longstanding issue in extragalactic formation history studies.
Gaia's 6D kinematic measurements in our Milky Way (MW) have vastly extended the possibilities for Galactic archaeology, leading to the discovery of a multitude of early mergers in the MW's past. As recent work has established a link between younger globular clusters (GCs; less than about 10--\SI{11}{\giga\year} old) and wet galaxy merger events, the MW provides an ideal laboratory for testing which GC properties can be used to trace extragalactic galaxy formation histories.}
{To test the hypothesis that GCs trace wet mergers, we relate the measured GC age distributions of the MW and three nearby galaxies, \m{31}, \ngc{1407}, and \ngc{3115}, to their merger histories and interpret the connection with wet mergers through an empirical model for GC formation.}
{The GC ages of observed galaxies are taken from a variety of studies to analyze their age distributions side-by-side with the model. For the MW, we additionally cross-match the GCs with their associated progenitor host galaxies to disentangle the connection to the GC age distribution. For the modeled GCs, we take galaxies with similar GC age distributions as observed to compare their accretion histories with those inferred through observations.}
{We find that the MW GC age distribution is bimodal, mainly caused by younger GCs (10--\SI{11}{\giga\year} old associated with Gaia-Sausage/Enceladus (GSE) and in part by unassociated high-energy GCs. The GSE GC age distribution also appears to be bimodal. We propose that the older GSE GCs (12--\SI{13}{\giga\year} old) were accreted together with GSE, while the younger ones formed as a result of the merger.
For the nearby galaxies, we find that clear peaks in the GC age distributions coincide with active early gas-rich merger phases. Even small signatures in the GC age distributions agree well with the expected wet formation histories of the galaxies inferred through other observed tracers.
From the models, we predict that the involved cold gas mass can be estimated from the number of GCs found in the formation burst.
}
{Multimodal GC age distributions can trace massive wet mergers as a result of GCs being formed through them.
From the laboratory of our own MW and nearby galaxies we conclude that the ages of younger GC populations of galaxies can be used to infer the wet merger history of a galaxy.}


\maketitle
%

\section{Introduction}
\label{sec:introduction}

Over the course of galaxy formation and evolution, in situ formed structures mix with accreted matter, concealing their origin and thereby the formation history of the galaxy. Through precise measurements of star and gas properties, such as their distribution in space, their velocity, their chemical compositions, and the stellar ages, it is possible to disentangle many of the individual clues on the formation history. This is the aim of galaxy archaeology \citep[e.g.,][]{freeman&bland_hawthorn02,binney13,helmi20}.
In particular, stellar structures and overdensities, such as stellar streams and other tidal features, can reveal details on a galaxy's history, where especially stellar streams are generally the remains of a tidally disrupted satellite galaxy. This has been done extensively for the Milky Way (MW; e.g., \citealp{helmi+99,belokurov+06,belokurov+07,bell+08,shipp+18}), aided particularly by the Gaia mission \citep{gaia_collaboration16:dr1_summary,gaia_collaboration+18:dr2_summary,gaia_collaboration+23:dr3_summary} in recent years with 6D phase-space data \citep[e.g.,][]{helmi+18,helmi20,prudil+22:II,malhan+22,ruiz_lara+22}.
It is also possible to detect tidal features for other galaxies in a more limited way through photometric and integral field unit (IFU) observations. Such identified structures have also been connected to the merger history of their host galaxies in observations \citep[e.g.,][]{bilek+20,bilek+23,chandra+23} and simulations \citep[e.g.,][]{bullock&johnston05,johnston+08,amorisco15,hendel&johnston15,pop+18,karademir+19,valenzuela&remus24}. However, all of these are tracers for stellar-dominated merger events and cannot trace the gas that has been accreted through such mergers.

For our own Galaxy, data are available in unprecedented detail, including measured proper motions. By detecting kinematic substructures of stars clustered in phase space, it has been possible to identify a number of past and ongoing mergers for the MW. The \object{Sagittarius Dwarf Spheroidal Galaxy} was the first merger discovered with the MW through positional and line-of-sight kinematic data by \citet{ibata+94}.
Using Hipparcos data and line-of-sight distances and velocities, \citet{helmi+99} discovered substructures in the inner halo now known as the Helmi streams.
Through the Gaia mission and the 6D kinematic data that were obtained for stars in the MW, a large number of halo stars were found to have distinct kinematics in phase space, which were attributed to a major merger event that is expected to have taken place around \SI{10}{\giga\year} ago, Gaia-Sausage/Enceladus (GSE; \citealp{belokurov+18,haywood+18,helmi+18,mackereth+19}). It is the last major merger that the MW experienced, as well as the most massive one, forming a large part of the stellar halo. 
There are also other interpretations of the kinematic signatures, however, that the local halo cannot be dominated by GSE alone, but is composed of the remains of multiple mergers over a prolonged time \citep[e.g.,][]{donlon+20,donlon+22,donlon&newberg23}.
Finally, \citet{myeong+19} find a second group of halo stars kinematically and chemically different from GSE stars with overall retrograde motions, which they attribute to a separate merger event, referred to as Sequoia.
Further groups of stars in phase space have been found, which are likely the debris of disrupted galaxies that fell into the MW, though the connection to specific merger events is not yet clear (see \citealp{dodd+23} and \citealp{horta+23} for a current overview of structures found in phase space).

Through Gaia data, globular clusters (GCs) in the MW have also been used to further disentangle the Galactic formation history. Based on their 6D phase-space properties and their age-metallicity relations, some studies have linked the GCs with their likely progenitor host galaxies \citep{myeong+18,massari+19,forbes20,horta+20,callingham+22,belokurov&kravtsov24,chen&gnedin24:MWassembly}, such as to the MW itself as in situ GCs, or to some of the inferred accreted galaxies. These studies also reveal unassociated groups of GCs with low and high orbital energies, where it is proposed that a group of unassociated low-energy GCs are part of a further past merger event \citep{massari+19,forbes20,callingham+22}. Similar structures in phase space with overlapping properties are identified through different methods \citep[e.g.,][]{kruijssen+19:kraken,forbes20,horta+21,horta+23}.
However, it has also been shown that GCs can migrate in phase space over time, potentially making it difficult to disentangle the origin of GCs based on their phase-space properties alone \citep{pagnini+23}.
An overview of the positions on the sky and in phase space of the streams and GCs detected in the MW can be found in the figures of \citet{riley&strigari20}, \citet{malhan+22}, and \citet{mateu23:galstreams}.

Because of their intrinsic brightness, GCs have also been used as tracer populations in the outskirts of other galaxies to study their mass distribution, kinematics, and formation history \citep[e.g.,][]{peng+04:GCsII,foster+11,coccato+13,pota+13,zhu+14,dolfi+21,versic+24}.
Through their old age, GCs in particular have experienced a large part of a galaxy's history, making them valuable tracers for past merger events. However, the formation of GCs themselves is still poorly understood. For this reason, models and simulations have been developed to help constrain the details of their formation process. Highly-resolved hydrodynamic simulations help study the resolved formation of individual GCs \citep[e.g.,][]{kravtsov&gnedin05,lahen+19,mckenzie&bekki21}, and sub-grid models for GCs applied to isolated or cosmological simulations allow one to follow GC properties and their spatial distribution through time, making comparisons with observations of nearby galaxies possible \citep[e.g.,][]{bekki+05,kruijssen+11,pfeffer+18:emosaics,chen&gnedin22,de_lucia+24:GCs,doppel+23,reina_campos+23,chen&gnedin24:MW_M31}. Finally, empirical and semi-analytic GC formation models applied to cosmological merger trees of galaxies enable one to test the parameter space of a limited number of free parameters for a large number of galaxies \citep[e.g.,][]{beasley+02,choksi+18,el_badry+19,valenzuela+21,chen&gnedin23}. Such models can lead to a better understanding of the statistical properties of GC formation that are necessary to reproduce GC properties and relations as they are observed today \citep[e.g.,][]{spitler&forbes09,harris+15,harris+17,forbes+18,burkert&forbes20}.

The stars of GCs can be individually observed in the MW, such that reasonably good measurements of their ages can be determined through the color-magnitude diagram (CMD), which has been done for various GC subsamples in the MW \citep[e.g.,][]{salaris&weiss98,dotter+08,marin_franch+09:acsgcVII}. For extragalactic GCs, the ages have much larger uncertainties associated with them because only the integrated GC properties can be measured. For this reason, stellar population models and evolutionary tracks have to be used for extragalactic systems to determine the ages. Ages have been estimated from photometric studies of the integrated light of extragalactic GCs \citep[e.g.,][]{chies_santos+11,georgiev+12,de_brito_silva+22}, a technique which will also become increasingly important with the upcoming generation observatories such as Euclid, the Nancy Grace Roman Space Telescope, and the Vera C.\ Rubin Observatory \citep[e.g.,][]{lancon+21,dage+23,usher+23}. Even for integrated spectroscopic measurements of extragalactic GCs, there are biases towards younger ages compared to the CMD method. This bias arises from hot horizontal branch (HB) stars, whose presence is degenerate with younger stars \citep[e.g.,][]{worthey94,de_freitas_pacheco&barbuy95,beasley+02:LMC,cabrera_ziri&conroy22} and even causes issues when using state-of-the-art models \citep[e.g.,][]{perina+11,goncalves+20}. This has been done for GCs in galaxies in the Local Group \citep[e.g.,][]{beasley+05:II,schiavon+13:m31clustersV,wang+21} and for selected nearby galaxies \citep[e.g.,][]{usher+19}.

In this work, we used a recent empirical GC formation model with two formation pathways \citep[][the first pathway forms GCs in small halos, the second forms GCs in gas-rich mergers]{valenzuela+21,valenzuela23} to study in what way its second pathway of forming GCs in gas-rich wet mergers \citep[e.g.,][]{ashman&zepf92} can help shed light on the formation history of the MW and other nearby galaxies. The model has been shown to agree well with the observed numbers of GCs in galaxies from dwarf to galaxy cluster masses, where a linear relation has been found to exist between the number of GCs and the dark matter (DM) halo virial mass \citep[e.g.,][]{blakeslee+97,harris+17,forbes+18}, as well as with GC age distributions of the MW and nearby galaxies. We introduce the GC model and observational data in \cref{sec:data_method}. In \cref{sec:analysis}, we present a bimodal feature found in the observed GC age distribution of the MW and how it could be related to the predictions of the GC model. We then test and discuss these predictions in detail for the MW in \cref{sec:discussion_mw} and for other nearby galaxies in \cref{sec:discussion_nearby_galaxies}. Finally, we summarize and conclude the results in \cref{sec:conclusion}.

\section{Data \& method}
\label{sec:data_method}

In the following, the empirical GC model and the GC observational data used in this work are presented. The main property studied is the GC age distribution of galaxies.

\subsection{Globular cluster model}
\label{sec:model}

In this work, we used the empirical GC formation model introduced by \citet{valenzuela+21}, which builds on previous models and investigations by \citet{boylan_kolchin17}, \citet{choksi+18}, and \citet{burkert&forbes20}. The model employs two formation pathways for GCs: The first, the small halo pathway, forms GCs in small haloes as soon as a halo's virial mass surpasses a given threshold value, $M_\mathrm{seedGC}$. With equal probability, 0, 1, or 2 GCs are formed. The second, the wet merger pathway, is the formation pathway introduced by \citet{choksi+18} and triggers GC formation when the relative halo mass accretion rate surpasses a given threshold value, $A_\mathrm{min}$. The formed number of GCs is then determined by converting the available cold gas mass $M_\mathrm{gas}$ to a total GC mass via a conversion factor, $\eta_\mathrm{GC}$ \citep{kravtsov&gnedin05,li&gnedin14,choksi+18,valenzuela+21}:
\begin{equation}
    M_\mathrm{GC} = \num{1.8e-4} \eta_\mathrm{GC} M_\mathrm{gas}.
\end{equation}
By assuming a cluster initial mass function of
\begin{equation}
    \frac{dN}{dM} \propto M^{-2},
\end{equation}
the expected number of formed GCs is obtained as (combining eqs.~3, 6, and~7 of \citealp{valenzuela+21})
\begin{equation}
    \label{eq:ngc_wet_formed}
    \langle N \rangle = \exp \bigg( W \Big( \frac{\num{1.8e-4} \eta_\mathrm{GC} M_\mathrm{gas}}{M_\mathrm{min}} \Big) \bigg) - 1,
\end{equation}
where $W$ is the Lambert $W$ function, $\eta_\mathrm{GC} = 0.5$ for the best-fitting model, and $M_\mathrm{min} = \SI{e5}{\Msun}$ is the minimum mass that a GC needs at formation time to survive for a few \si{\giga\year} \citep{li&gnedin14,choksi+18}.
For more details on the models, see \citet{valenzuela+21}.
Note that recent work by \citet{chen&gnedin23} now uses a smaller value of $M_\mathrm{min} = \SI{e4}{\Msun}$, although they note that GCs with low initial masses of for example \SI{e4}{\Msun} have an estimated lifetime of around \SI{1}{\giga\year} at \SI{3}{\kilo\parsec} distance of the center of a MW-mass galaxy. The model only tracks the numbers of GCs per galaxy and their formation times, but does not include metallicities. This limits the comparison with observations to only the GC ages, though the available measured metallicities can be used as indicators for the formation sites of the observed GCs. In contrast, for the modeled GCs this information is already known.

The GC model was applied to the merger tree of a DM-only simulation of side length \SI{30}{\mega\parsec} with a DM particle mass of $m_\mathrm{DM} = \SI{7.90e6}{\Msun}$ that was run with the TreePM code \gadget{3} \citep{springel05:gadget2}. The empirical model \name{emerge} \citep{moster+18:emerge} provides the model with the baryonic matter content per galaxy.
Because the model only tracks the number of GCs in each galaxy and at what times they were formed, the model parameters were fit to match observations of the numbers of GCs, since those are available for a sufficiently large sample: the GC numbers are taken from \citet{burkert&forbes20}. The GC age distributions are consistent with those found by \citet{usher+19} for the MW and three SLUGGS galaxies (SAGES Legacy Unifying Globulars and GalaxieS; \citealp{brodie+14:sluggs}), and the fractions of GCs formed through the wet merger pathway agree with the red GC fractions measured by \citet{harris+15}. For more information on the comparability to observations and how the model parameters affect the resulting GC properties, see \citet{valenzuela+21}.

\subsection{Globular cluster observational data}
\label{sec:data}

A variety of observed GC age measurements from the literature are used in this work for selected galaxies in the Local Universe. These use different methods to obtain the GC ages and are presented in the following.
For all of the measured samples, it is important to keep in mind that while the absolute ages have large uncertainties and are difficult to measure even in the MW itself \citep[e.g.,][for \m{92}]{ying+23}, within a given GC sample they are subjected to the same systematic uncertainties, resulting in much more precise relative ages.
This is important for the study of GC age distributions, where the features in the distribution itself are given by the relative ages as opposed to the absolute ones.

\subsubsection{Milky Way}
\label{sec:mw_data}

The MW is the only galaxy besides its own satellites for which it is currently possible to obtain accurate CMDs of its GCs. This allows for a much more exact determination of their ages and has been done in multiple studies for different sized samples of GCs. In this work, we considered three of these as compiled by \citet{kruijssen+19:kraken}, and additionally investigated the mean ages of those three studies in \cref{app:all_gcs}:
\begin{itemize}
    \item \citet{forbes&bridges10} with a sample of 92~GCs,
    \item \citet{dotter+10:acsgcIX,dotter+11} with a sample of 68~GCs,
    \item \citet{vandenberg+13} with a sample of 54~GCs,
    \item \citet{kruijssen+19:kraken} with a sample of 96~GCs, which contains the mean GC ages from the previous three studies.
\end{itemize}

The sample from \citet{forbes&bridges10} is based on a number of previous age measurement studies \citep{salaris&weiss98,bellazzini+02,catelan+02,de_angeli+05,carraro+07,dotter+08,carraro09,marin_franch+09:acsgcVII}, of which 64~GCs were measured using the Advanced Camera for Surveys (ACS) from the Hubble Space Telescope (HST) through the ACS survey for Galactic GCs \citep{sarajedini+07:acsgcI,marin_franch+09:acsgcVII} to obtain relative ages. They were normalized to absolute ages with the Dartmouth models of \citet{dotter+07:acsgcII}. While that sample is restricted to the inner \SI{20}{\kilo\parsec} of the MW, the age measurements of further GCs were supplemented from the other works.
The sample from \citet{dotter+10:acsgcIX,dotter+11} is for the most part also based on the GCs measured through the ACS survey of Galactic GCs using the photometric catalog from \citet{anderson+08:acsgcV}, and the remaining GCs were observed with further HST/AST measurements.
Lastly, the GC age measurements from \citet{vandenberg+13} were computed from the same photometric catalog of the ACS survey of Galactic GCs as \citet{marin_franch+09:acsgcVII} used, but employing the stellar evolutionary tracks from \citet{vandenberg+12}.
It should be noted that while not all these objects are necessarily actual GCs, but in part also nuclear star clusters, metal complex clusters, or a combination thereof \citep[e.g.,][]{mckenzie+22}, we refer to all of them simply as GCs in this work.

In addition to the three CMD age samples of the MW GCs, we also included two GC age samples obtained through integrated measurements, as this is also what one is restricted to for other galaxies:
\begin{itemize}
    \item \citet{usher+19} with a sample of 61~GCs. Their measurements are obtained through combining photometry and spectroscopy, to which stellar population models are fitted using a Markov chain Monte Carlo (MCMC) method.
    \item \citet{cabrera_ziri&conroy22} with a sample of 32~GCs, of which we removed the 3~spurious young GCs (see their section~6.1, where they detail how the spectral fit residuals indicate whether the best fit for a young GC is real or spurious, which is a result of the simple single population HB star model that they use). These have also been shown to be much older from CMD measurements.
    Compared to \citet{usher+19}, \citet{cabrera_ziri&conroy22} used spectroscopy only, but additionally took the hot horizontal branch (HB) stars into account in their modeling of the integrated stellar population measurements.
\end{itemize}

For the MW, additional GC properties can be determined that are not possible to obtain for other galaxies at the moment.
Six-dimensional phase space measurements have been made available for many MW GCs through Gaia \citep{gaia_collaboration18:dr2_gckinematics,vasiliev19}, which \citet{massari+19} used to assign the likely origin of the individual GCs. Their list of progenitors for the GCs consist of the MW itself (i.e., in situ formed GCs in the disk or bulge), the GSE galaxy, the Sagittarius dwarf, the Helmi Streams, the Sequoia galaxy, and unassociated high- and low-energy GCs. \Citet{forbes20} used the age-metallicity relation (AMR) to improve these progenitor assignments and propose that the unassociated low-energy GCs belong to a single progenitor dwarf satellite, which they dub Koala and is likely related to or overlaps with Kraken \citep{kruijssen+19:kraken} and Heracles \citep{horta+21,horta+23}. Further work was done by \citet{callingham+22}, who used a chemo-dynamical model and hydrodynamical simulations of MW-like galaxies to associate the GCs with their progenitor hosts. Their assumed accretion events largely align with those used by \citet{massari+19} and \citet{forbes20}.
\Citet{chen&gnedin24:MWassembly} have recently applied clustering techniques to chemical, spatial, kinematic, and age properties of the GCs to associate the GCs with the MW, GSE, Sagittarius, and other ex-situ-formed GCs.
It should be noted that due to the ongoing observations and work on this topic, this is a rapidly evolving field.
The identifiers of the GCs were available to us for the CMD age samples and for the sample of \citet{cabrera_ziri&conroy22}, such that we could cross-match the ages for those four samples to the progenitor assignments. For this work, we use the assignments made by \citet{forbes20}, though using the assignments from \citet{massari+19}, \citet{callingham+22}, or \citet{chen&gnedin24:MWassembly} do not change the statistical findings presented in this work when we did the analysis using those instead (also see \cref{app:gc_associations_alternatives} for an analysis using the associations from \citealp{callingham+22} and \citealp{chen&gnedin24:MWassembly}).

\subsubsection{\texorpdfstring{\m{31}}{M 31}}
\label{sec:m31_data}

As the nearest more massive galaxy to the MW, the Andromeda galaxy (\m{31}) is an ideal galaxy for which GCs can be identified and analyzed, since observations have much better resolution and better signal-to-noise values than for more distant galaxies. For \m{31}, we used two different studies for the GC ages:
\begin{itemize}
    \item \citet{wang+21} with a sample of 343~clusters, of which we used the 293 old GCs ($t_\mathrm{age} > \SI{1.5}{\giga\year}$) in this work,
    \item \citet{usher+24} with a sample of 290~clusters,
    \item Cabrera-Ziri et al.\ (in prep.) with a sample of 286~GCs, of which we used the 136~GCs whose ages are sufficiently constrained.
\end{itemize}

The sample from \citet{wang+21} was observed with the Large Sky Area Multi-Object Fiber Spectroscopic Telescope (LAMOST; \citealp{cui+12:lamost,luo+15:lamost_dr1}). The GC ages were then determined based on the obtained integrated spectra and multi-band photometry from Beijing-Arizona-Taiwan-Connecticut (BATC; \citealp{ma+15}) and Sloan Digital Sky Survey (SDSS; \citealp{peacock+10}). For the old clusters, they fit the spectra using an empirical stellar spectra library and the measured colors using an MCMC method. Their ages are all younger than \SI{10}{\giga\year}, however, which is likely a consequence of their underlying models and almost certainly not actually the case for the majority of \m{31}'s GCs \citep[e.g.,][]{beasley+05:II,caldwell+11:m31clustersII,schiavon+13:m31clustersV}. Their ages should therefore be taken with caution.
The GC age determinations from \citet{usher+24} were also made from photometric and spectroscopic measurements, where the photometry was obtained with SDSS \citep[mostly from][]{peacock+10} or from the Pan-Andromeda Archaeological Survey (PAndAS) Canada France Hawaii Telescope MegaCam for the halo GCs not covered by SDSS \citep{huxor+14}. \Citet{usher+24} used the same method as \citet{usher+19}, for which they fit the stellar spectra using an MCMC method.

The integrated spectra measurements of Cabrera-Ziri et al.\ (in prep.) use the same method as applied for the MW GCs presented by \citet{cabrera_ziri&conroy22}. Their GC sample is selected from the inner halo of \m{31}, of which 150 have only a lower bound on their age, leaving 136~GCs with sufficiently constrained ages to study their distribution. The reason for the large number of lower age bounds is that their method is able to determine that the metal-poor GCs with a horizontal branch are very old, but not what exact age they have (${>}\SI{9}{\giga\year}$ in all cases, and ${>}\SI{12.5}{\giga\year}$ is the median lower bound) due to degeneracies between the ages and horizontal branch properties. This also means that the resulting age distribution has a selection bias towards metal-rich GCs, which means that the metal-poor GCs typically accreted through smaller galaxies are removed. This should be kept in mind, though we believe that the consequences for this work are not severe since we focus on GCs formed during mergers, which produce younger GCs at generally higher metallicities. GCs formed through such mergers tend to be younger by on average 2--\SI{3}{\giga\year}, having their formation peak at an age of roughly \SI{10}{\giga\year} \citep[see fig.\ 2 of][]{valenzuela23}.

\subsubsection{\texorpdfstring{\ngc{3115}}{NGC 3115} and \texorpdfstring{\ngc{1407}}{NGC 1407}}
\label{sec:sluggs_data}

The age measurements for 116~GCs in \ngc{3115} and 213~GCs in \ngc{1407} were made by \citet{usher+19} by combining photometry and spectroscopy to fit the stellar spectra using an MCMC method. The spectral data were obtained through the SLUGGS survey \citep{brodie+14:sluggs} and the photometric data are from \citet{arnold+11} using the Suprime-Cam of the Subaru telescope.

\section{Age dating wet mergers with GCs in the Milky Way}
\label{sec:analysis}

Having the observational data, we investigated some reoccurring features in the GC age distributions with the aim of verifying if such features can be explained by the model. In the following, we present our initial findings from the observations and what predictions the model makes with respect to these.

\subsection{GC age distribution observations in the Milky Way}
\label{sec:observed_distributions}

\begin{figure}
    \centering
    \includegraphics[width=\columnwidth]{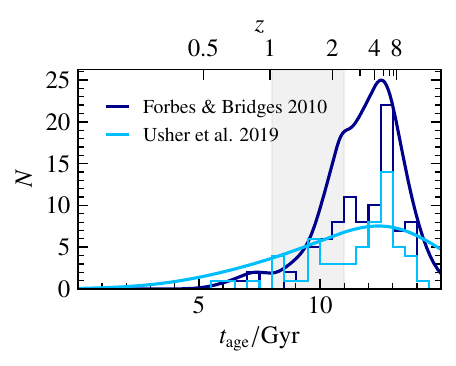}
    \caption{Globular cluster age distributions in the MW from \citet{forbes&bridges10} and \citet{usher+19}, with sample sizes of 92 and 61~GCs, respectively. The boxy lines are the actual histograms, while the smooth curves show the distributions smoothed by the measurement uncertainties. These are computed through a summation over normal distributions with the respective ages as the means and their uncertainties as the standard deviations. The shaded region between \SIlist{8;11}{\giga\year} indicates the estimated time of the GSE merger \citep{belokurov+18,helmi+18}.}
    \label{fig:mw_gc_intro}
\end{figure}

As the galaxy for which the best data exists, we first considered the GC age distribution of the MW. In two of the available samples, the ages appear to have a bimodal or even multimodal distribution, which are shown in \cref{fig:mw_gc_intro}. For both \citet{forbes&bridges10} and \citet{usher+19}, the main peak of the distribution is at \SI{13}{\giga\year}. There is also a slight second peak in the distribution of \citet{forbes&bridges10} at around \SI{11}{\giga\year}, while for \citet{usher+19} there are two small peaks between \SIlist{8;10}{\giga\year}.
Since the data from \citet{forbes&bridges10} are composed of multiple different studies, we show the age distribution only of the largest of the underlying GC age studies from \citet{marin_franch+09:acsgcVII} in \cref{app:marinfranch_gcs} (it includes 64 of the 92~GCs), which also features the age bimodality. Thus, the bimodality seen for \citet{forbes&bridges10} is not the result of combining multiple studies with different systematic uncertainties.
Studies of the age-metallicity relationship for MW GCs have shown that there are multiple branches of GCs corresponding to the in situ formed GCs and different progenitor galaxies that fell into the MW \citep[e.g.,][]{kruijssen+19:kraken,forbes20}, which could be what is seen as a multimodal age distribution. The peaks in the age distributions align well with the estimated infall times of Sagittarius (8--\SI{9}{\giga\year} ago) and Sequoia (${\sim}\SI{10}{\giga\year}$ ago; \citealp{forbes20}), and of the GSE merger event, the last major merger that the MW experienced \citep{helmi+18,haywood+18}: around \SIrange{8}{11}{\giga\year} ago according to \citet{belokurov+18} and around \SI{10}{\giga\year} ago according to \citet{helmi+18}. With an estimated merger mass ratio of 1:4 \citep{helmi+18,gallart+19}, GSE is expected to have been a gas-rich major merger that triggered a starburst in the disk of the MW \citep[e.g.,][]{helmi20,ciuca+23}. There is also evidence for this from stellar population measurements, where \citet{gallart+19} found a clear peak of high star formation at around \SI{9.5}{\giga\year} ago.

Of course, the findings from the observed age distributions are accompanied by some caveats: First of all, GC age measurements are subject to large uncertainties, especially for the older ages (see the solid lines in \cref{fig:mw_gc_intro}, which are the distributions smoothed by Gaussian kernels given by the measurement uncertainties). Still, a large part of these uncertainties are systematic effects, such that the relative ages can still be trusted more than individual absolute ages. Second, the sample sizes are not large enough to make any kind of statistically significant statements, in particular for the few GCs that make up the minor peaks in the age distributions. Third, bringing together the estimated time of the GSE merger event and the time of the GC age distribution peaks is far from having established a causal relation between the two. However, since the features are present in the data and the time also aligns with that of the GSE merger event, GC formation models can be employed to investigate if there is a theoretical basis for a causal connection.

\subsection{Model predictions}
\label{sec:predictions}

\begin{figure}
    \centering
    \includegraphics[width=0.97\columnwidth]{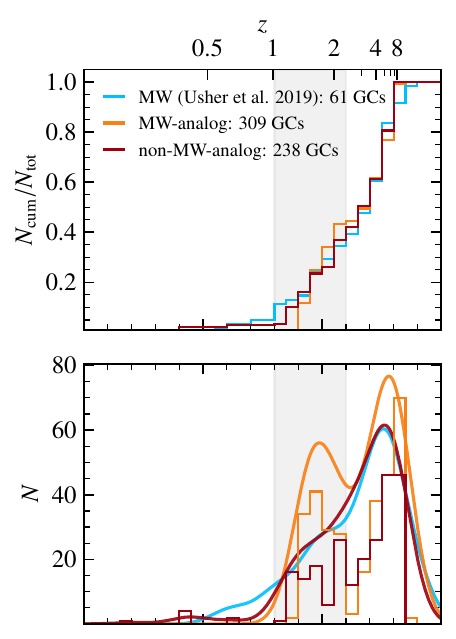}
    \includegraphics[width=0.96\columnwidth]{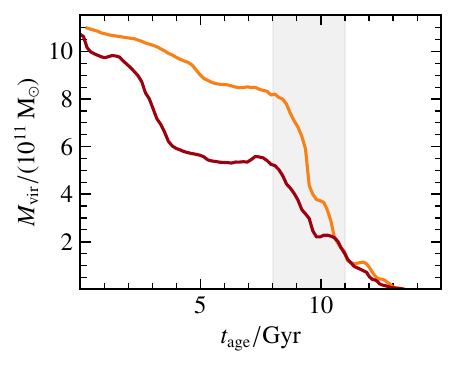}
    \caption{GC ages and virial mass evolutions of two selected modeled MW-mass galaxies compared to the observed GC ages in the MW.
    \emph{Top}: Cumulative GC age distributions of the MW from \citet{usher+19} (blue) and of two modeled GC populations in simulated MW-mass galaxies from \citet{valenzuela+21} (red and orange).
    \emph{Middle}: Age distributions of the same three GC populations as in the top panel. The smooth distributions are smoothed using an assumed uncertainty of \SI{0.75}{\giga\year} for the GC ages. These are computed through a summation over normal distributions with the respective ages as the means and their uncertainties as the standard deviations. For the sample from \citet{usher+19}, the distribution was scaled by a factor of four to be comparable to the modeled populations.
    \emph{Bottom}: Dark matter virial mass evolution of the two modeled galaxies, showing their accretion histories. The shaded region between \SIlist{8;11}{\giga\year} indicates the estimated time of the GSE merger \citep{belokurov+18,helmi+18}.}
    \label{fig:mw_gc_analogs}
\end{figure}

The model introduced by \citet{valenzuela+21} consists of two GC formation pathways, of which one allows GCs to form through wet mergers containing a sufficient amount of cold gas. To study what the model predicts for galaxies such as the MW, we first extracted MW-like galaxies from the simulation to which the GC model was applied. For this, we selected MW-mass galaxies by their virial mass of $\Mvir = 1$--\SI{2e12}{\Mvir}, which applies to 21 simulated galaxies.
Out of these, there are two with GC age distributions that best match the distribution measured by \citet{usher+19} for the MW in terms of their cumulative distributions, which is shown in the top panel of \cref{fig:mw_gc_analogs}. The analogs were selected using the measurements by \citet{usher+19} instead of one of the CMD measurements since the model was originally calibrated by \citealp{valenzuela+21} to be in agreement with the ages measured by \citealp{usher+19} to have a comparison with multiple galaxies. A recalibration of the model to the CMD age distributions is unsuitable because it would surpass the scope of this work and would only provide one single galaxy with a sufficient number of CMD-measured GC ages.
The conclusions drawn in the following are still applicable to the model in general, independent of the age calibration that was used. We show the GC age distributions and the accretion histories of the other 19 MW-mass galaxies in \cref{app:other_mw_like_galaxies}.

Both of the simulated galaxies have roughly bimodal GC age distributions with a minor peak at around \SIrange{9}{10}{\giga\year} (middle panel of \cref{fig:mw_gc_analogs}). Assuming a measurement uncertainty of \SI{0.75}{\giga\year} for each GC age continues to show a clear bimodality in one case (orange distribution), but only leaves a hint for the other case (red distribution). This shows that even if there is a bimodal age distribution, the large measurement uncertainties for GC ages can make it difficult to actually confirm it in practice.

The underlying reason for these peaks in the age distributions becomes apparent when studying the accretion histories of the two galaxies: both of them experience a major merger in the same time period as the GC formation bursts (bottom panel of \cref{fig:mw_gc_analogs}). The mergers occur at the same time as the GSE merger is estimated to have happened (8--\SI{11}{\giga\year} ago). The two accretion histories differ strongly in their later evolution, however: while the galaxy with the more pronounced bimodal GC age distribution only experiences mini to minor mergers afterwards (orange line in the bottom panel of \cref{fig:mw_gc_analogs}), the other galaxy has a second major merger at a later time of around 2--\SI{4}{\giga\year} ago (red line). The lack of GCs having formed around that time is a clear indication that the merger was rather gas-poor (i.e., dry). The formation history of the first galaxy is therefore more similar to that inferred for the MW (\enquote{MW-analog}), while the other galaxy has had a much more violent recent history (\enquote{non-MW-analog}). The fact that the non-MW-analog GC age distribution seen in the middle panel of \cref{fig:mw_gc_analogs} is more similar to the observed one by \citet{usher+19} only indicates that the major merger around \SI{10}{\giga\year} ago is more similar to the GSE merger in terms of their GCs formed. However, the second major merger around 2--\SI{4}{\giga\year} ago is in no way comparable to the MW, making it the non-MW-analog.
As shown for the other 19 MW-mass galaxies in \cref{app:other_mw_like_galaxies}, the peaks in their GC age distributions also correspond to gas-rich merger events, while those mergers that do not leave a strong imprint in the GC age distribution are the dry mergers that the galaxy experienced.

In conclusion, the model makes the following predictions: wet mergers with a large amount of cold gas are capable of producing a bimodal or even multimodal GC age distribution for a galaxy, providing an indication for a type of event that is generally difficult to trace through other means. However, dry mergers with little to no cold gas do not result in noticeable signatures in the GC age distributions. This is the case for the majority of late-time mergers, but at higher redshifts gas-rich mergers are increasingly more common and beyond $z \approx 2$ the most common kind of merger event \citep[e.g.,][]{bournaud+11}. Thus, GC ages provide a means to probe the very early turbulent formation times of galaxies by tracing their massive wet merger events.

\section{Discussion: GCs in the Milky Way}
\label{sec:discussion_mw}

Having the model prediction that wet mergers leave an imprint on the GC age distribution of a galaxy, we tested it through the available GC measurements in the MW. For this we could make use of additional properties currently unavailable for GCs around other galaxies, such as kinematic phase-space information and more accurate age measurements.

\subsection{Milky Way diagnostics}
\label{sec:mw_diagnostics}

\begin{figure*}
    \centering
    \includegraphics[width=\textwidth]{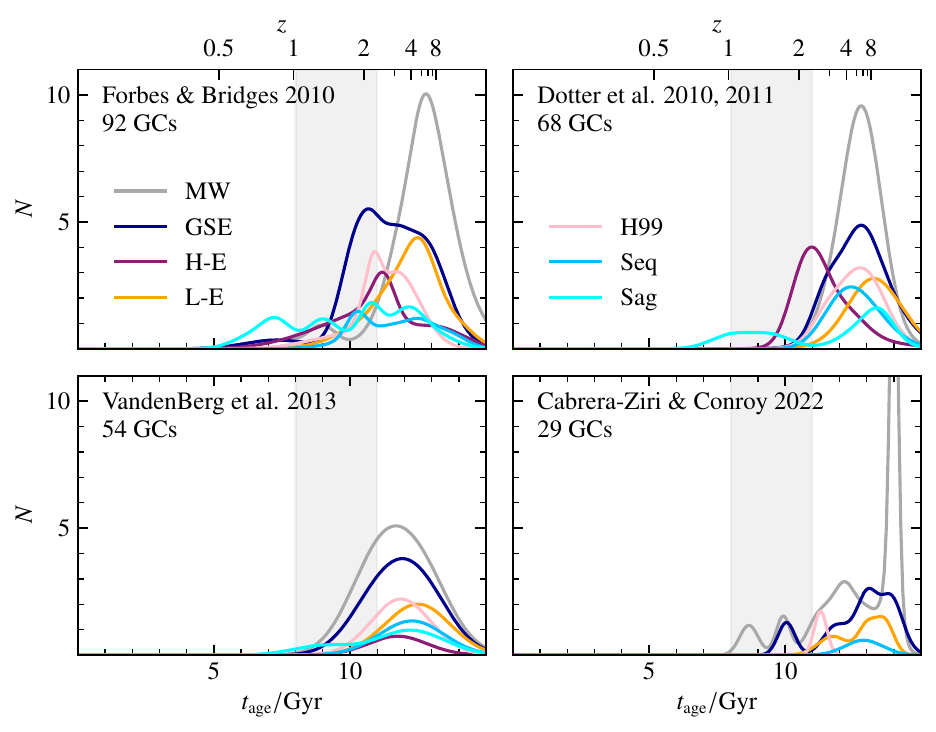}
    \caption{Globular cluster age distributions in the MW from \citet{forbes&bridges10}, \citet{dotter+10:acsgcIX,dotter+11}, \citet{vandenberg+13}, and \citet{cabrera_ziri&conroy22}, split up according to their likely progenitors from \citet{forbes20}. The shown classifications are the MW, GSE, unassociated high-energy GCs (H-E), unassociated low-energy GCs (L-E, which was given the name Koala by \citealp{forbes20}), the Helmi Streams (H99), Sequoia (Seq), and Sagittarius (Sag). The total number of GCs in the respective sample is indicated in the top left. The distributions are computed from the GC ages and their uncertainties through a summation over normal distributions with the respective ages as the means and their uncertainties as the standard deviations. The shaded region between \SIlist{8;11}{\giga\year} indicates the estimated time of the GSE merger \citep{belokurov+18,helmi+18}. See \cref{fig:mw_gc_panel_hist} for the corresponding histograms of the most relevant GC progenitor groups without taking the uncertainties into account.}
    \label{fig:mw_gc_panel_smooth}
\end{figure*}

To address some of the caveats brought up in \cref{sec:observed_distributions}, we used further data available on the MW GCs to study to what extent the model prediction can be applied to the MW and its GC population. Using the GC progenitor assignments by \citet{forbes20}, the age distributions for four of the samples \citep{forbes&bridges10,dotter+10:acsgcIX,dotter+11,vandenberg+13,cabrera_ziri&conroy22} can be split up by those assignments. \Cref{fig:mw_gc_panel_smooth} shows the age distributions for the GCs formed in situ in the MW and those associated with GSE and the other GC host progenitors, for each of the four samples. This figure uses the age uncertainties to smooth the distributions.
For the total GC age distributions see \cref{fig:mw_gc_all}, which shows that for the smoothed distributions, only the total age distribution from \citet{forbes&bridges10} shows a hint at a bimodality. This bimodality is smoothed out when taking the mean GC ages from \citet{kruijssen+19:kraken}, which may lead to GC age biases (discussed in \cref{app:all_gcs}).
See \cref{app:gc_unsmoothed_histograms} for the unsmoothed age distributions plotted as histograms for the most relevant GC host progenitor groups that correspond to \cref{fig:mw_gc_panel_smooth}.

The two smaller samples from \citet{vandenberg+13} and \citet{cabrera_ziri&conroy22} do not reveal further information besides an indication for the low-energy GCs to have formed at slightly later times for the measurements by \citet{vandenberg+13}. In contrast, both of the other samples show that there is a physical reason for the bimodal GC age distribution shown in \cref{fig:mw_gc_intro}: in the largest sample by \citet{forbes&bridges10}, it is clearly seen that the younger peak seen in \cref{fig:mw_gc_intro} at \SI{11}{\giga\year} is dominated by GSE-associated GCs, with contributions from the Helmi Streams and the high-energy GCs. This is less pronounced in the sample by \citet{dotter+10:acsgcIX,dotter+11}, where the high-energy GCs contribute most of the young GCs, albeit there is also a significant contribution from GSE. The fact that these GCs have a different origin than the other MW GCs is known through the age-metallicity relation \citep[e.g.,][]{leaman+13,kruijssen+19:kraken,forbes20}, in which the GCs associated with GSE and the other progenitor galaxies follow their own tracks.
While the Sagittarius dwarf is expected to have fallen into the MW around 8--\SI{9}{\giga\year} ago on a most likely rather circular orbit, its extended time range of GC formation could be a result of tidally induced star cluster formation during its orbit around the MW while there was still enough cold gas available \citep[e.g.,][for a study on young clusters with ages of around \SI{2}{\giga\year} in the \object{Small Magellanic Cloud}, which are suggested to have formed through tidal interactions with the \object{Large Magellanic Cloud}]{williams+22}.

It is curious, however, that there appears to be a hint at a non-unimodal age distribution within the GSE GCs, which can be seen more clearly for \citet{forbes&bridges10} than for \citet{dotter+10:acsgcIX,dotter+11} in \cref{fig:mw_gc_panel_smooth}. The bimodality is also present when only using the GC ages from the subsample of \citet{marin_franch+09:acsgcVII} (\cref{app:marinfranch_gcs}) as well as when using the mean GC ages from \citet{kruijssen+19:kraken}. The bimodality is even slightly more prominent when using those ages, despite the overall distribution being more smoothed out (\cref{app:all_gcs}). Additionally, it is also present when using the GC progenitor associations from any of \citet{massari+19}, \citet{callingham+22}, or \citet{chen&gnedin24:MWassembly}, and it is actually much more clearly visible for their classifications (\cref{app:gc_associations_alternatives}).
The GCs associated with GSE are therefore shown to have an age bimodality across multiple different studies of their ages and association models with the GSE.
The reason the bimodality is not seen in the samples from \citet{dotter+10:acsgcIX,dotter+11} and \citet{vandenberg+13} is that they each only contain 13~GCs associated with GSE, compared to 21~GCs from \citet{forbes&bridges10}.
Considering the raw age data without uncertainties, both distributions are actually bimodal (\cref{fig:mw_gc_panel_hist}), with the caveat of there being a very small amount of GCs involved.

Still, if the signal is real, it can be brought together with the model prediction in a straightforward manner.
As the model forms GCs in small halos at early times (first pathway as described in \cref{sec:model}), a GSE satellite galaxy hosts its own GCs as it falls into the MW. We refer to these GCs as the accreted GCs in the following. Assuming there is enough cold gas available, the major merger event then triggers GC formation through the gas collision and tidal forces between the two galaxies (merger-induced GCs in the following, second pathway as described in \cref{sec:model}; \citealp{ashman&zepf92,williams+22}). This scenario leads to multiple properties arising for the GCs: (1) The age distribution of the combined early-formed accreted and merger-induced GCs is bimodal. (2) Assuming that GSE brings in its own cold gas, many of the merger-induced GCs are also formed from that gas, or a mixture of that and the MW's gas and therefore have similar metallicities and phase-space properties as those of the accreted GCs. This would then also lead to such GCs being associated with GSE through an analysis of phase-space and the age-metallicity relation. (3) Assuming the merger leads to violent gas interactions, it is possible that some recently formed merger-induced GCs are ejected from the overall orbit of GSE, leading to unassociated high-energy GCs. In that case, the high-energy GCs could also be associated with GSE and thus the bimodality in the GSE GC age distribution would be even more enhanced. It is also possible for GCs to distribute themselves further apart in phase space through other dynamical processes as shown by \citet{pagnini+23}, potentially ending up in the high-energy regime. This may occur in a similar fashion to the Splash \citep[e.g.,][]{bonaca+17,belokurov+20}, a group of more metal-rich stars in the MW halo that appear to have been formed in situ and dynamically ejected around the time of the GSE merger \citep[e.g.,][]{belokurov+20,ciuca+23}. Of course, this would not result in a change in age and metallicity of the formed GCs, though \citet{ciuca+23} argue that the GSE merger could have first driven down the metallicity, which was then again enriched by the induced starburst.

While it is not possible to prove this theory with the current state of GC age precision and the difficulties of kinematic associations, the GC formation model indicates that the GCs associated with GSE are not only those that were brought in by the accreted galaxy \citep[e.g.,][]{cote+98}, but also those that were formed through the wet merger \citep[e.g.,][]{ashman&zepf92}. Additionally, it is possible that some of the unassociated high-energy GCs were also formed in the process of the GSE merger, though it should be noted that many of those GCs have lower measured metallicities than those associated with GSE \citep{forbes20}. In turn, this could be a result of the metallicity lowering through the merger \citep{ciuca+23}. However, analyzing and modeling the details of GC metallicities in such a gas-rich major merger scenario are beyond the scope of this work and will be addressed in a future study.

\subsection{Model diagnostics}
\label{sec:model_diagnostics}

From \cref{eq:ngc_wet_formed} it is possible to obtain the number of GCs expected from the model to form through a merger event based on the cold gas mass, $M_\mathrm{gas}$.
\Citet{vincenzo+19} estimated GSE to have brought in a cold gas mass of \SI{6.62e9}{\Msun}. For the MW, we estimated a range of possible cold gas masses by determining the cold gas masses of galaxies at $z=2$ in a hydrodynamical cosmological simulation since the simulation that our GC model was applied to only contains DM particles. For this, we used Magneticum Pathfinder\footnote{\url{www.magneticum.org}} Box4 (uhr) (with a side length of \SI{68}{\mega\parsec} and a gas particle mass of $m_\mathrm{gas} = \SI{1.0e7}{\Msun}$) because it has the necessary resolution to resolve MW-like galaxies and their formation well, and at the same time it is large enough to contain a large sample of MW-like galaxies, from which we can draw conclusions about the expected gas mass at $z=2$. The galaxies contained within it have been shown to agree well with observations across a broad range of properties (see \citealp{teklu+15} for details on the implementations and \citealp{valenzuela&remus24} for an overview of comparisons to observations).
For the galaxies at $z=2$ with virial masses $\SI{5e11}{\Msun} \leq M_\mathrm{vir} \leq \SI{8e11}{\Msun}$, the typical cold gas mass (which we define as star-forming gas particles with temperatures below \SI{e5}{\kelvin}) is $M_\mathrm{gas} = \SI{2.1\pm0.6e10}{\Msun}$.
We computed the expected numbers of formed GCs for a range of different MW cold gas masses between \SIlist{1.0e10;7.5e10}{\Msun}, leading to 6 to 22 GCs being formed (\cref{tab:predicted_ge_gcs}).

\begin{table}
    \centering
    \caption{Predicted number of GCs, $N_\mathrm{GC,formed}$, formed through the GSE merger with the MW for different assumed MW total cold gas masses, $M_\mathrm{gas,MW}$. The GSE galaxy is assumed to have had a cold gas mass of \SI{6.62e9}{\Msun} \citep{vincenzo+19}.}
    \label{tab:predicted_ge_gcs}
    \begin{tabular}{cc}
        \hline\hline
        $M_\mathrm{gas,MW} / \si{\Msun}$ & $N_\mathrm{GC,formed}$ \\
        \hline
        \num{1.0e10} & $6$ \\
        \num{2.5e10} & $11$ \\
        \num{5.0e10} & $17$ \\
        \num{7.5e10} & $22$ \\
        \hline
    \end{tabular}
\end{table}

As seen in the GC age distribution from \citet{forbes&bridges10}, there are 12 out of 21~GCs with available ages associated with GSE and 7 out of 9~unassociated high-energy GCs in the time range of GSE and the secondary peak of the overall MW GC age distribution. While the latter GCs are surely not all related to the GSE merger, the GC sample also does not include all GCs found in the MW (around 50--\SI{60}{\percent}). Overall, the number of GCs expected from the model to be formed through such a merger aligns well with the observed number of GCs found to be associated with GSE or to be unassociated with high energies, while also taking into account the statistical incompleteness of the GC sample. This supports the prediction that multimodal GC age distributions can be used to trace wet mergers.

\section{Discussion: Extension to nearby galaxies}
\label{sec:discussion_nearby_galaxies}

\begin{figure}[!ht]
    \centering
    \includegraphics[width=\columnwidth]{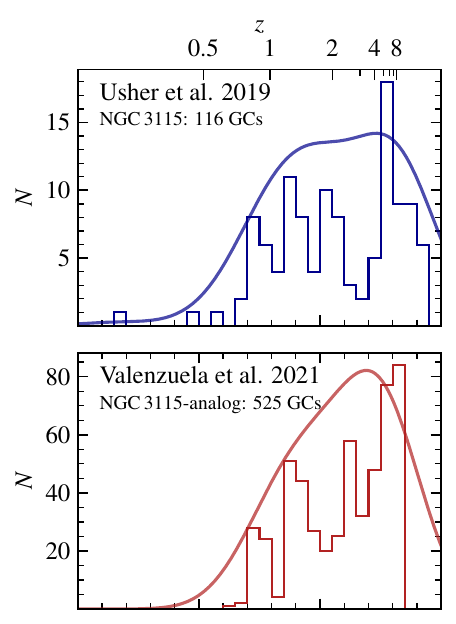}
    \includegraphics[width=\columnwidth]{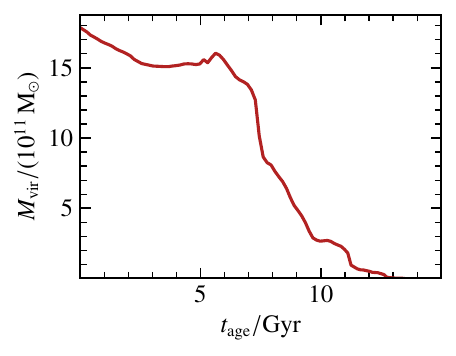}
    \caption{GC ages and virial mass evolution of the modeled \ngc{3115}-analog galaxy compared to the observed GC ages in \ngc{3115}.
    \emph{Top and middle}: Age distributions of the GC population in \ngc{3115} from \citet{usher+19}, with a sample of 116~GCs, and of the modeled GC population in an \ngc{3115}-analog galaxy from \citet{valenzuela+21} with 525~GCs. The smooth distributions are smoothed using an assumed uncertainty of \SI{1.5}{\giga\year} for the GC ages. These are computed through a summation over normal distributions with the respective ages as the means and their uncertainties as the standard deviations. \emph{Bottom}: Dark matter virial mass evolution of the simulated \ngc{3115}-analog galaxy, showing its accretion history.}
    \label{fig:ngc3115_gc}
\end{figure}

While the GC age measurements are by far the most accurate for the MW due to resolved measurements of the clusters, the general properties from integrated GC age measurements can still give indications about the wet merger history of the host galaxy, through the relative ages between the GCs. One of the galaxies studied by \citet{usher+19} is \ngc{3115}, a fast-rotating S0 field galaxy with stellar mass $M_* = \SI{9e10}{\Msun}$ and virial mass $\Mvir = \SI{1.2e12}{\Msun}$ (\citealp{forbes+16}, assuming an NFW profile such that the virial mass is a factor 10 larger than the DM mass within $8\,R_e$, where $R_e$ is the effective radius, the radius within which half the light of the galaxy is emitted). It has \num{550\pm80}~GCs \citep{harris+13} and features multimodal age distributions (based on 116~measured GC ages from \citealp{usher+19}; top panel of \cref{fig:ngc3115_gc}). The multimodal behavior of the ages is even retained when smoothing the histogram with a Gaussian kernel of \SI{1.5}{\giga\year} (smooth line), which is a value we selected to illustrate the effect of smoothing that can be caused by measurement uncertainties. There is in fact a galaxy in the simulation that has a very similar GC age distribution as \ngc{3115}, which can here be seen in the middle panel of \cref{fig:ngc3115_gc} (the similarity of the distributions is seen especially well in the cumulative distributions shown in fig.~17 of \citealp{valenzuela+21}). The simulated galaxy has a virial mass of $\Mvir = \SI{1.8e12}{\Msun}$, similar to that of \ngc{3115}, and a total of 525~GCs, consistent with \ngc{3115}, which places it \SI{0.2}{\dex} above the mean linear scaling relation from \citet{burkert&forbes20}, but still within the observed scatter. Interestingly, both the observed and simulated ages feature three minor peaks in the distribution without smoothing over it, although this could very well be a coincidence given the large uncertainties for the observational measurements. For the simulation this means that there were overall three especially gas-rich mergers leading to an increased amount of GC formation. This occurred during a time of steady assembly between \SIlist{6;12}{\giga\year} ago (bottom panel of \cref{fig:ngc3115_gc}).

This agrees well with the inferred formation history of \ngc{3115} obtained through kinematics from IFU data and tracer populations, the metallicity profiles, its mass distribution, and the study of its morphological components: it is believed to have experienced an early gas-rich accreting phase followed by a lack of significant mergers thereafter \citep{arnold+11,guerou+16,zanatta+18,poci+19,dolfi+20,buzzo+21}. In particular, this supports the prediction of GC ages being able to trace wet mergers, also for galaxies outside the Local Group.

Such behavior is not the norm, however. This could already be seen from the median age distributions presented by \citet{valenzuela+21} for all the virial mass bins, which show that generally GC populations are dominated by old GCs like those of the MW. For most galaxies, GC formation bursts are too close in time to the oldest GC populations to be able to distinguish them without having further properties available like in the MW, or the bursts are not significant enough due to a lack of cold gas as gas-rich merger events become less and less frequent with time.

\begin{figure}
    \centering
    \includegraphics[width=\columnwidth]{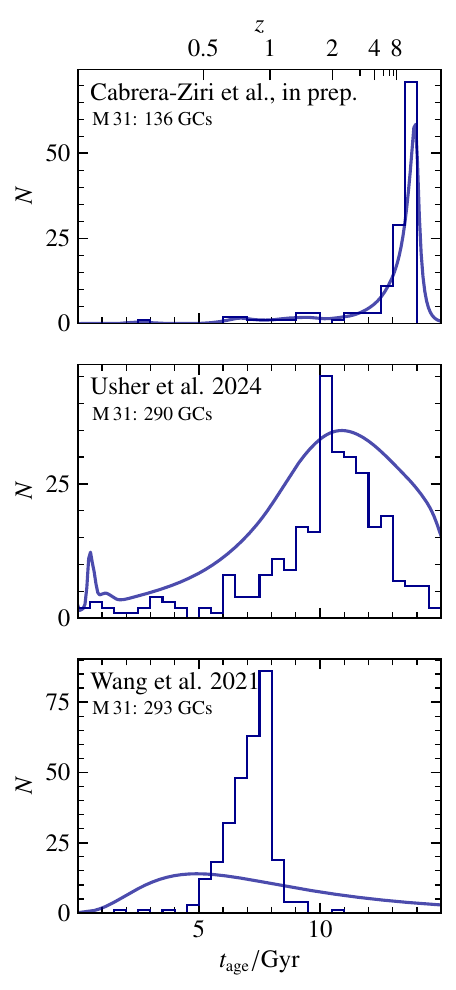}
    \caption{Globular cluster age distribution in \m{31} from Cabrera-Ziri et al.\ (in prep.), \citet{usher+24}, and \citet{wang+21}, with sample sizes of 136~GCs, 290~GCs, and 293~GCs, respectively. The ages in the sample of \citet{wang+21} correspond to their determined old clusters ($t_\mathrm{age} > \SI{1.5}{\giga\year}$). For further comparisons with previous GC age determinations in \m{31}, see fig.~8 of \citet{wang+21}. The smooth lines show the distributions smoothed by the measurement uncertainties. These are computed through a summation over normal distributions with the respective ages as the means and their uncertainties as the standard deviations for \citet{usher+24} and over log-normal distributions for the other two data sets. Note that for the logarithmic uncertainties, the visualization in linear space is skewed with respect to the peak.}
    \label{fig:m31_gc}
\end{figure}

One such case is \m{31}, for which the GC age measurements of \citet{wang+21}, \citet{usher+24}, and Cabrera-Ziri et al.\ (in prep.) show no clear bimodal distribution (\cref{fig:m31_gc}). There is a hint at some additional modes between \SI{6}{\giga\year} and \SI{10}{\giga\year} ago in the data from \citet{usher+24} and Cabrera-Ziri et al.\ (in prep.) (middle and lower panel). The numbers of GCs in these modes are small for the latter study (6 around 6--\SI{8}{\giga\year} ago and 8 around 8--\SI{10}{\giga\year} ago) and the uncertainties of the GC age measurements are large compared to the individual age peaks (smooth curves in the middle and lower panel), so we cannot exclude that they originate from statistical uncertainties. If they do not, this would indicate only small gas-rich mergers given the small numbers of GCs produced.
As discussed in \cref{sec:m31_data}, the absolute ages of the measured distributions should not be compared with those of \citet{wang+21} due to differences in the measurement techniques, which led to the much younger determined ages for \citet{wang+21}. In terms of their relative ages, however, all of the samples show that significant features in the GC age distributions are not found for all galaxies and strong features are only visible for mergers that have a sufficient amount of gas. Additionally, which features are identifiable in GC age distributions will always depend on how large the measurement errors are, which smooth out the data, which first removes the signatures from smaller mergers or those with smaller gas fractions. Thus, the method is the most reliable in detecting wet mergers with large mass fractions. Finally, note that the sample from Cabrera-Ziri et al.\ (in prep.) is biased towards metal-rich GCs (\cref{sec:m31_data}). However, gas-rich major mergers are expected to involve the more metal-rich GCs as opposed to metal-poor ones, such that we believe the implications of our analysis to be unaffected.

Observationally, it has been proposed that \m{31} experienced a major merger at around \SI{2}{\giga\year} ago, possibly with the progenitor of \m{32} \citep{dsouza&bell18:m31,dsouza&bell18:mergers,hammer+18}. A large peak in star formation between \SIlist{2;4}{\giga\year} ago in the disk and outskirts of \m{31} \citep{bernard+12,bernard+15,williams+17} is likely related to this event. Using planetary nebulae, \citet{bhattacharya+23:VI} find further evidence for a wet major merger of \m{31} \SI{2.5}{\giga\year} to \SI{4}{\giga\year} ago. In fact, the full sample by \citet{wang+21} also includes young stellar clusters with ages below \SI{1.5}{\giga\year}, which could potentially also have been formed as a result of the gas brought in by the merger and would not be referred to as GCs yet. Similarly, the distribution by \citet{usher+24} shows GCs with these kind of young ages as well as GCs with ages around 3--\SI{4}{\giga\year}, and Cabrera-Ziri et al.\ (in prep.) also find one GC with an age of \SI{2.5}{\giga\year}, which could coincide with that merger.

It has also been determined that the star formation rate was very low before the recent peak, with most stars having formed prior to \SI{8}{\giga\year} ago \citep{williams+17}. Due to the smooth and very massive stellar halo observed for \m{31}, it is estimated that the merger history was dominated by many smaller accretion events \citep[e.g.,][]{ibata+14,mackey+19}. This scenario is supported by the model prediction presented in this work, in which no single sufficiently massive merger exists that would lead to a significant GC formation burst.

\begin{figure}
    \centering
    \includegraphics[width=\columnwidth]{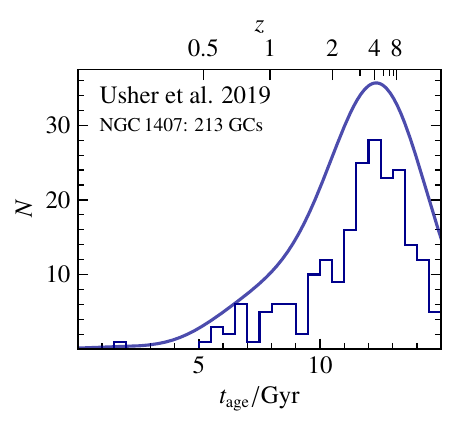}
    \caption{Globular cluster age distribution in \ngc{1407} from \citet{usher+19} with a sample size of 213~GCs. The smooth line shows the distribution smoothed using an assumed uncertainty of \SI{1.5}{\giga\year} for the GC ages. These are computed through a summation over normal distributions with the respective ages as the means and their uncertainties as the standard deviations.}
    \label{fig:ngc1407_gc}
\end{figure}

Finally, the massive elliptical galaxy \ngc{1407} also lacks a significant second peak in its GC age distribution, though there is a tail with a slight peak towards younger ages around 6--\SI{9}{\giga\year} ago and potentially another around \SI{10}{\giga\year} ago (\cref{fig:ngc1407_gc}).
Due to the integrated age measurements, this could be the result of underestimated ages for those GCs as the peak again consists of only very few GCs. Overall, it appears that from the GC ages \ngc{1407} has not experienced any massive wet mergers since the early buildup phase of the galaxy, and at most a later merger with a low cold gas fraction. In the latter case, it should be expected that there would also be a sign of late star formation activity in the stellar populations themselves. In fact, \citet{spolaor+08} found from their stellar population measurements of \ngc{1407} that the stars are uniformly old, having formed around \SI{12}{\giga\year} ago.

A kinematically decoupled core (KDC) hints at a major merger of \ngc{1407} with gas fractions between \SI{15}{\percent} and \SI{40}{\percent} \citep{hoffman+10,schulze+17}. While \citet{forbes&remus18} find that a simulated galaxy from Magneticum with a similar size and mass as \ngc{1407} had a late major merger around \SI{8}{\giga\year} ago, it was not selected based on further properties such as stellar ages or metallicity gradients. However, \citet{ferre_mateu+19} find in their observations of \ngc{1407} that the KDC is slightly younger than the rest of the galaxy (we estimate a difference of around \SI{1}{\giga\year} based on their fig.~5), suggesting that a wet major merger could have occurred slightly later than the early buildup of the galaxy. This scenario is compatible with the GC age distribution found by \citet{usher+19}: the lack of late gas-rich mergers is consistent with no large amount of late GC formation, and the slight peak in the GC ages could be related to the wet major merger that formed the KDC. Such a merger is expected to have had a relatively high gas fraction \citep{hoffman+10}, potentially forming many GCs as a result. However, due to the violent nature of such a merger, many of those systems would likely be disrupted in the same process, leading to a smaller peak in the age distribution. Since the GC sample analyzed by \citet{usher+19} is biased towards central GCs due to the SLUGGS survey having been focused on the inner regions, it is expected that it would be more likely to pick up signatures from major mergers.

\section{Conclusion}
\label{sec:conclusion}

In this work, we have used the GC formation model from \citet{valenzuela+21} with dual formation pathways to predict that massive wet mergers with enough cold gas can leave an imprint on the age distribution of the GC population in the host galaxy. This imprint results in a bimodal or even multimodal distribution, indicating when the wet mergers occurred. This prediction is in part also a consequence of the idea that red GCs tend to form through mergers that also induce star formation, thus resulting in properties that overall trace the underlying stellar component, such as spatial, kinematic, or chemical properties \citep[e.g.,][]{brodie&strader06,pota+13,dolfi+21}. In contrast, mergers with little to no gas are not traced by the GC ages since the lack of gas means that no significant number of GCs could be formed in the process.

The prediction is discussed for the MW in detail. We find that a hint at bimodality visible in the pure GC age distributions compiled by \citet{forbes&bridges10} can be further disentangled by combining the data with phase-space information on the GCs to map which galaxy progenitors the GCs are likely associated with \citep[e.g.,][]{massari+19,forbes20,callingham+22,chen&gnedin24:MWassembly}. We find the later peak in the GC age distribution to correspond to the GCs associated with GSE, the last major merger of the MW \citep{belokurov+18,helmi+18}, and in part also to the unassociated high-energy GCs and the GCs of the Helmi Streams. In fact, the age distribution of the GSE GCs appears to also have a bimodality. Since GSE is expected to have been a massive and gas-rich merger, we suggest that GSE not only brought in its own older GCs, but also formed a second group of GCs through the merger with the MW. The second group of GCs would be located near the first in phase space and lie on the same age-metallicity relation, as is also found for them in the observations. We further suggest that some of the unassociated high-energy GCs may also originate from the GSE merger, since it is possible that the violent merger dynamics would eject some GCs, or that GCs migrate away in phase space over time \citep{bonaca+17,belokurov+20,pagnini+23}.

Two simulated MW-mass galaxies with similar GC age distributions as the MW are found to both have had a wet major merger around the same time as GSE (around \SI{8}{\giga\year} to \SI{10}{\giga\year} ago). One of the two then evolved only through smooth accretion and mini and minor mergers, as is believed to have been the case for the MW. In contrast, the other simulated galaxy encountered a dry merger at later times that is not traced by the GC ages, (for those kind of mergers, other kind of indicators have to be used).

We also tested the model with three other observed galaxies, \ngc{3115}, \m{31}, and \ngc{1407}, for which GC ages have been obtained. For \ngc{3115}, the observed GC age distribution is clearly multimodal with a considerable population of younger GCs that are 7--\SI{11}{\giga\year} old.
We were able to find a simulated galaxy of comparable virial mass and a similar GC age distribution. It underwent an early phase of multiple wet mini and minor mergers that led to the formation of the younger GC population, after which it experienced no further significant mergers. This agrees well with the expected formation history of \ngc{3115} inferred from other tracers \citep{arnold+11,guerou+16,poci+19,buzzo+21} and supports the prediction that additional modes in the GC age distribution trace wet mergers.

The GC age distribution of \m{31} features no bimodality, indicating that it experienced no significant late wet mergers. However, it shows two or even three small peaks in the GC age distributions at 2--\SI{3}{\giga\year} and at 7~and \SI{9}{\giga\year}, indicating either small merger events with large gas fractions or large merger events with low gas fractions, with the former being more likely than the latter. Since these peaks are small and the uncertainties are large, this is not conclusive, however. Similarly, \ngc{1407} features no clear bimodality, but a possible second late peak in the GC age distribution may be related to an early wet major merger that led to the kinematically distinct core found in the overall very old galaxy. These are examples for the most common case for galaxies: the simulation predicts that strong peaks in the GC age distribution with GCs younger than around \SI{12}{\giga\year} clearly indicate a massive wet merger event and that small wet mergers can still leave minor peaks in the GC age distribution. Dry mergers, however, leave no imprints in the GC age distributions.

To conclude, the age distribution of GCs can be used as a tracer for wet mergers of galaxies, which are generally more difficult to infer from observations: while the old age peak of the GC age distribution does not help in constraining the merger history as here the old populations from the main progenitor of a galaxy mix with the old GCs brought in through other merging galaxies of all kind, the young GCs with ages less than around \SI{11}{\giga\year} are formed in the otherwise untraceable wet merger events. However, due to the current large observational uncertainties in determining GC ages from integrated measurements, further development of the age determination techniques will be essential to better understand individual galaxies' formation histories. Finally, increasing the number of galaxies with accurate GC age measurements will also help improve our understanding of GC formation and set more constraints on current GC formation models.
We thus propose that this is a good method to infer the wet merger times of extragalactic galaxies from observations.
With the advent of observatories such as Euclid, the Nancy Grace Roman Space Telescope, and the Vera C.\ Rubin Observatory, applying photometric methods to obtain GC ages from such observations has the great potential of providing deeper insight into the early gas-rich formation history of galaxies.

\begin{acknowledgements}
We thank Ivan Cabrera-Ziri and Giulia Pagnini for helpful discussions.
We thank the anonymous referee for useful comments that improved the manuscript.
LMV acknowledges support by the German Academic Scholarship Foundation (Studienstiftung des deutschen Volkes), the Marianne-Plehn-Program of the Elite Network of Bavaria, and the COMPLEX project from the European Research Council (ERC) under the European Union’s Horizon 2020 research and innovation program grant agreement ERC-2019-AdG 882679.

This research was supported by the Excellence Cluster ORIGINS, funded by the Deutsche Forschungsgemeinschaft under Germany's Excellence Strategy – EXC-2094-390783311.

The following software was used for this work: \code{astropy} \citep{astropy_collaboration13:astropy,astropy_collaboration18:astropy}, \code{jupyter} \citep{kluyver+16:jupyter}, \code{matplotlib} \citep{hunter07:matplotlib}, \code{numpy} \citep{harris+20:numpy}, \code{pandas} \citep{mckinney10:pandas, pandas:pandas}, Julia \citep{bezanson+17:julia}, CSV.jl \citep{quinn+:csv.jl}, and DataFrames.jl \citep{kaminski+:dataframes.jl}.
\end{acknowledgements}

\bibliographystyle{style/aa}
\bibliography{bib}

\begin{appendix}
    
\section{Milky Way globular cluster age distributions}

\subsection{Age distributions of all globular clusters}
\label{app:all_gcs}

\begin{figure}
    \centering
    \includegraphics[width=\columnwidth]{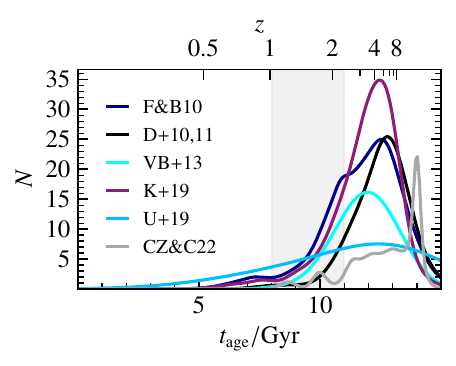}
    \caption{Globular cluster age distributions in the MW from \citet{forbes&bridges10}, \citet{dotter+10:acsgcIX,dotter+11}, \citet{vandenberg+13}, \citet{kruijssen+19:kraken}, \citet{usher+19}, and \citet{cabrera_ziri&conroy22}, smoothed by the measurement uncertainties. The curves are computed through a summation over normal distributions with the respective ages as the means and their uncertainties as the standard deviations. The shaded region between \SIlist{8;11}{\giga\year} indicates the estimated time of the GSE merger \citep{belokurov+18,helmi+18}.}
    \label{fig:mw_gc_all}
\end{figure}

The age distributions of the full GC samples of \citet{forbes&bridges10}, \citet{dotter+10:acsgcIX,dotter+11}, \citet{vandenberg+13}, \citet{kruijssen+19:kraken}, \citet{usher+19}, and \citet{cabrera_ziri&conroy22} are shown in \cref{fig:mw_gc_all}. The values from \citet{kruijssen+19:kraken} are obtained through the mean ages from the first three samples, all of which used CMD measurements, which leads to the total distribution being further smoothed out. Since not all of the CMD samples contain all 96~GCs from that study, the mean ages of the individual GCs will be biased towards younger or older ages depending on which samples they are contained in due to differences in the systematic uncertainties between the samples.

\begin{figure}
    \centering
    \includegraphics[width=\columnwidth]{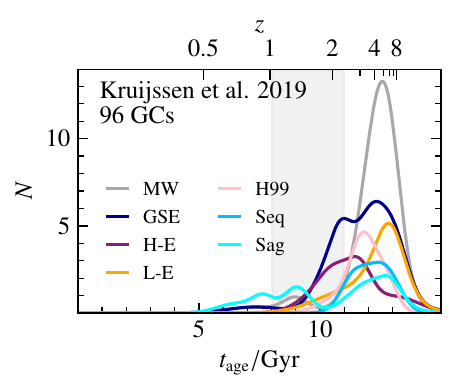}
    \caption{Globular cluster age distributions in the MW from \citet{kruijssen+19:kraken}, split up according to their likely progenitors from \citet{forbes20}. The shown classifications are the MW, GSE, unassociated high-energy GCs (H-E), unassociated low-energy GCs (L-E, which was given the name Koala by \citealp{forbes20}), the Helmi Streams (H99), Sequoia (Seq), and Sagittarius (Sag). The curves are smoothed by the measurement uncertainties, which are computed through a summation over normal distributions with the respective ages as the means and their uncertainties as the standard deviations. The shaded region between \SIlist{8;11}{\giga\year} indicates the estimated time of the GSE merger \citep{belokurov+18,helmi+18}.}
    \label{fig:mw_gc_mean}
\end{figure}

Nevertheless, the bimodality in the GSE GC age distribution is still clearly visible (\cref{fig:mw_gc_mean}), if not even clearer than in the top left panel of \cref{fig:mw_gc_panel_smooth} for the ages from \citet{forbes&bridges10}. This strongly indicates that the found bimodality is an actual feature and not a statistical artifact.

\subsection{Alternative progenitor associations}
\label{app:gc_associations_alternatives}

\begin{figure}
    \centering
    \includegraphics[width=\columnwidth]{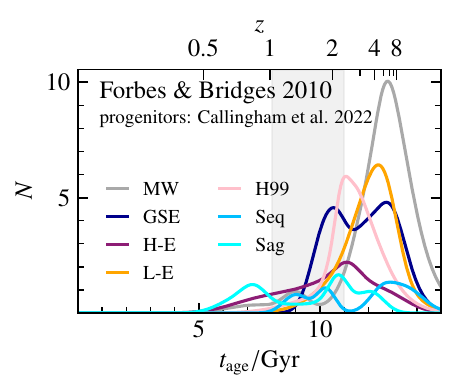}
    \caption{Globular cluster age distributions in the MW from \citet{forbes&bridges10}, split up according to their likely progenitors from \citet{callingham+22}. The shown classifications are the MW, GSE, unassociated high-energy GCs (H-E), unassociated low-energy GCs (L-E, which was given the name Koala by \citealp{forbes20}), the Helmi Streams (H99), Sequoia (Seq), and Sagittarius (Sag). The shaded region between \SIlist{8;11}{\giga\year} indicates the estimated time of the GSE merger \citep{belokurov+18,helmi+18}.}
    \label{fig:mw_gc_callingham}
\end{figure}

\begin{figure}
    \centering
    \includegraphics[width=\columnwidth]{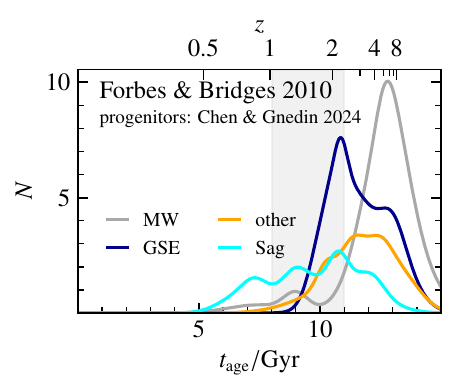}
    \caption{Globular cluster age distributions in the MW from \citet{forbes&bridges10}, split up according to their likely progenitors from \citet{chen&gnedin24:MWassembly}. The shown classifications are the MW, GSE, Sagittarius (Sag), and other ex-situ formed GCs (other). The shaded region between \SIlist{8;11}{\giga\year} indicates the estimated time of the GSE merger \citep{belokurov+18,helmi+18}.}
    \label{fig:mw_gc_chengnedin}
\end{figure}

While we used the GC progenitor associations from \citet{forbes20} in this work, it makes no qualitative difference to use the associations determined by other studies. For instance, the data from \citet{callingham+22} show a somewhat different distribution of GCs among GSE, the Helmi Streams, and the low-energy GCs, but the GC age distributions of GSE still features a bimodal distribution (\cref{fig:mw_gc_callingham}). This bimodality is visible in a much more distinct way when compared with the top left panel of \cref{fig:mw_gc_panel_smooth}, which shows the distributions for the GC associations determined by \citet{forbes20}.
Similarly, when using the associations from \citet{chen&gnedin24:MWassembly}, \cref{fig:mw_gc_chengnedin} shows that the age distributions change for GSE and Sagittarius (the only two ex-situ groups they specifically identify), but again the bimodality of the GSE-associated GCs persists. This time, the younger peak at \SI{11}{\giga\year} is significantly stronger than the ones seen from the associations of \citet{forbes20} and \citet{callingham+22}, which would mean that potentially more GCs were formed through the GSE merger.

\subsection{Age distributions of a single study}
\label{app:marinfranch_gcs}

\begin{figure}
    \centering
    \includegraphics[width=\columnwidth]{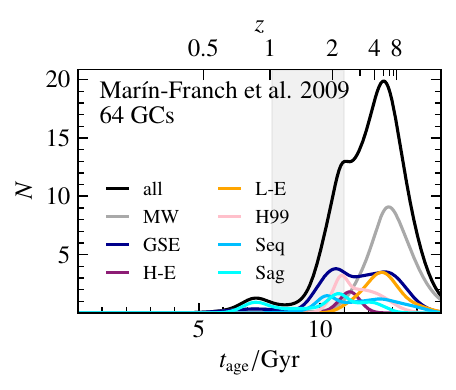}
    \caption{Globular cluster age distributions in the MW from \citet{marin_franch+09:acsgcVII}, split up according to their likely progenitors from \citet{forbes20}. The shown classifications are the MW, GSE, unassociated high-energy GCs (H-E), unassociated low-energy GCs (L-E, which was given the name Koala by \citealp{forbes20}), the Helmi Streams (H99), Sequoia (Seq), and Sagittarius (Sag). The curves are smoothed by the measurement uncertainties, which are computed through a summation over normal distributions with the respective ages as the means and their uncertainties as the standard deviations. The shaded region between \SIlist{8;11}{\giga\year} indicates the estimated time of the GSE merger \citep{belokurov+18,helmi+18}.}
    \label{fig:marinfranch_gcs}
\end{figure}

The GC sample of \citet{forbes&bridges10} is composed of multiple different studies (see \cref{sec:mw_data}), of which the largest subsample is of \citet{marin_franch+09:acsgcVII}, which includes 64 of the 92~GCs.
As combining several GC studies could cause problems due to different systematic biases, we show the single GC study by \citet{marin_franch+09:acsgcVII} in \cref{fig:marinfranch_gcs}.
Both the bimodality of the overall GC age distribution as well as in the GSE GC age distribution are clearly visible. This demonstrates that these bimodalities do not result from combining several GC studies, but in fact come from a single study.

\begin{figure*}
    \centering
    \includegraphics[width=\textwidth]{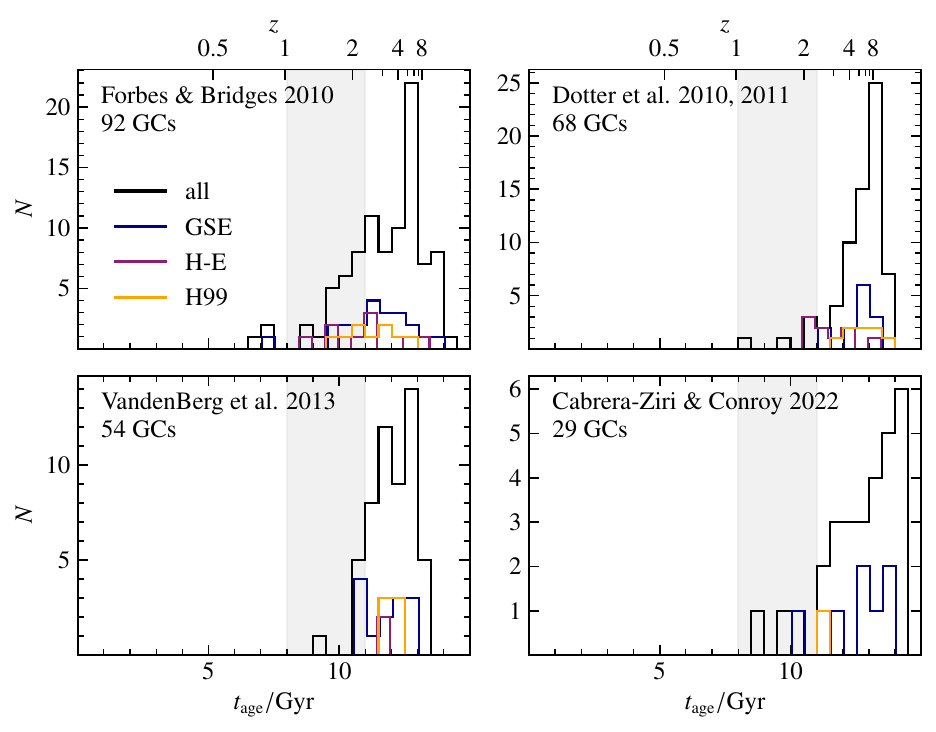}
    \caption{Globular cluster age distributions in the MW from \citet{forbes&bridges10}, \citet{dotter+10:acsgcIX,dotter+11}, \citet{vandenberg+13}, and \citet{cabrera_ziri&conroy22}, split up according to their likely progenitors from \citet{forbes20}. The shown classifications are the total population (all), GSE, the unassociated high-energy GCs (H-E), and the Helmi Streams (H99). The total number of GCs in the respective sample is indicated in the top left. The shaded region between \SIlist{8;11}{\giga\year} indicates the estimated time of the GSE merger \citep{belokurov+18,helmi+18}. See \cref{fig:mw_gc_panel_smooth} for the corresponding age distributions smoothing the data points through their uncertainties.}
    \label{fig:mw_gc_panel_hist}
\end{figure*}

\subsection{Age histograms}
\label{app:gc_unsmoothed_histograms}

The unsmoothed age distribution of the MW are shown in \cref{fig:mw_gc_panel_hist} for the four GC samples by \citet{forbes&bridges10}, \citet{dotter+10:acsgcIX,dotter+11}, \citet{vandenberg+13}, and \citet{cabrera_ziri&conroy22}, using the progenitor associations given by \citet{forbes20}. It becomes apparent that GSE, the unassociated high-energy GCs, and the GCs from the Helmi Streams are important contributors to the second peak in the total GC age distribution. The smoothed distributions according to the measured age uncertainties are found in \cref{fig:mw_gc_panel_smooth}.

\section{Other Milky Way-mass modeled globular cluster age distributions}
\label{app:other_mw_like_galaxies}

\begin{figure*}
    \centering
    \includegraphics[width=\textwidth]{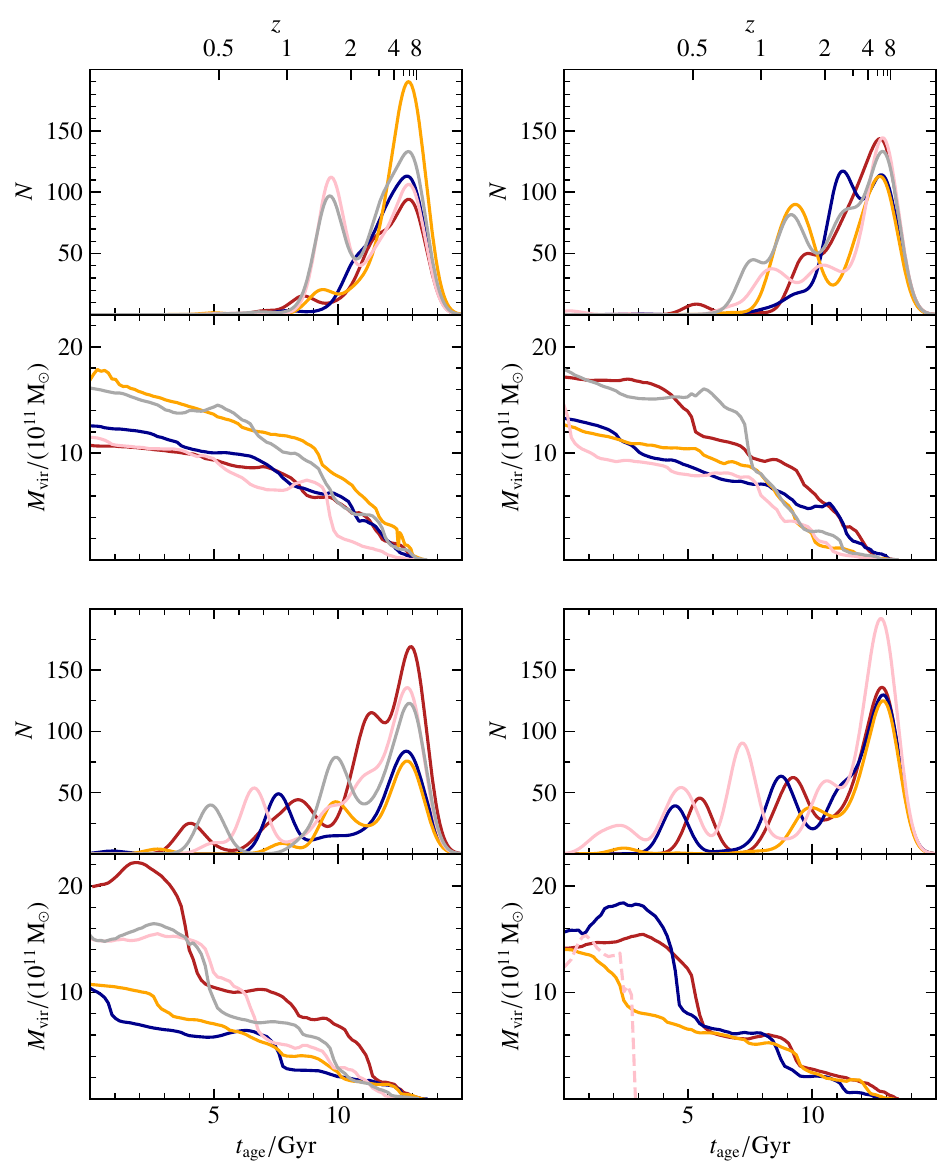}
    \caption{Age distributions of the globular clusters in 19 simulated MW-mass galaxies and their virial mass evolutions plotted below the respective age distributions in the same colors for each galaxy. The distributions are smoothed using an assumed uncertainty of \SI{0.5}{\giga\year} for the GC ages. The galaxies shown in the top left panels are the galaxies that reached half of their $z=0$ virial mass at the earliest time, followed by the galaxies in the top right panel. The galaxies in the bottom right panel are those that reach half of the current virial mass the latest. The galaxy with the dashed virial mass evolution line has an issue with the merger tree, but the GC population is correct.}
    \label{fig:other_mw_like_galaxies}
\end{figure*}

\begin{figure}
    \centering
    \includegraphics[width=\columnwidth]{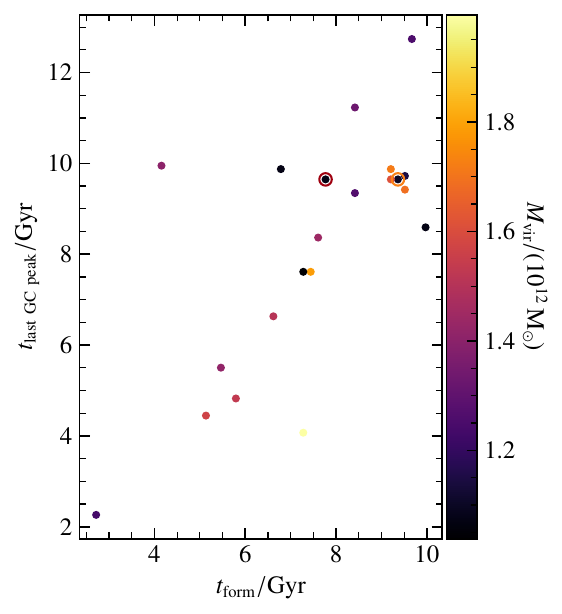}
    \caption{Correlation between the age of the youngest GC peak in their age distribution ($t_\text{last GC peak}$) and the formation time of the MW-mass simulated galaxies ($t_\mathrm{form}$), colored by the virial mass at $z=0$. The youngest peak is required to have an absolute value above~10. The formation time of a galaxy is taken as the lookback time at which the galaxy has reached half of its virial mass at $z=0$. The MW-analog and non-MW-analog from \cref{fig:mw_gc_analogs} are additionally circled in orange and red, respectively.}
    \label{fig:t_form_t_last_peak}
\end{figure}

As we found 21 MW-mass galaxies in the simulation with virial masses of $\Mvir = 1$--\SI{2e12}{\Mvir}, but only discussed the two of them with GC age distributions most similar to those observed in the MW (see \cref{sec:predictions}), we here present the other 19 simulated galaxies. \Cref{fig:other_mw_like_galaxies} shows their GC age distributions (top of each of the four double panels) and their respective virial mass evolutions in the same color (bottom of the double panels). From the top left to the bottom right, the galaxies are sorted by their formation time, that is at what time the galaxy reached half of the virial mass that it has at $z=0$, where the galaxies formed the earliest are shown at the top left. It can immediately be seen that the galaxies formed the earliest also have the oldest GC populations, whereas the later-formed galaxies have increasingly young GC subpopulations. The time of the youngest GC formation peak is furthermore strongly correlated with the formation time as seen in \cref{fig:t_form_t_last_peak}. From the MW-analog, we also obtain a prediction for the formation of the MW itself of around 9--\SI{10}{\giga\year}, which corresponds to $z=1.3$--1.7.
Overall, the ages of the youngest GCs also directly mark the time of the last major wet accretion event, as expected from the GC model.

Overall, the oldest GC population is almost always the dominant one (with the exception of the double-peaked age distributions in the upper left panel shown in pink and in the upper right panel shown in blue). In these two cases, there are nearly as many of the oldest GCs as the second youngest peak in the age distribution contains.

\end{appendix}

\end{document}